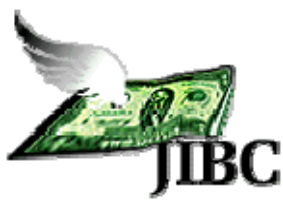
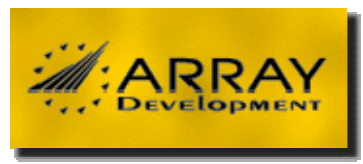

# PIXASTIC: STEGANOGRAPHY BASED ANTI-PHISHING BROWSER PLUG-IN

**P.Thiyagarajan,**
**Research Scholar, Department of Computer Science, Pondicherry University**
*Postal Address:* **CDBR-SSE Lab, Department of Computer Science, School of Engineering and Technology, Pondicherry University, Puducherry -14**.
Email: *thiyagu.phd@gmail.com*

Mr.P.Thiyagarajan is full time Ph.D Scholar in Department of Computer Science Pondicherry University. He obtained his Integrated M.Sc Computer Science with distinction from College of Engineering Guindy Anna University Main Campus Chennai. Prior to joining research he has couple of years experience in Software Industry. His research interest includes Information security, Network Security and Database Management System.

**G. Aghila,**
**Professor, CDBR-SSE Lab, Department of Computer Science, Pondicherry University**
*Postal Address:* **CDBR-SSE Lab, Department of Computer Science, School of Engineering and Technology, Pondicherry University, Puducherry -14**.
Email: *aghilaa@gmail.com*

Prof. G. Aghila is currently working in the Department of Computer Science, Pondicherry University. She obtained her undergraduate degree (B.E (CSE)) from TCE, Madurai in the year 1988. Her postgraduate degree M.E. (CSE) is from the School of Computer Science, Anna University, Chennai in the year 1991. She obtained her Doctorate in the field of Computer Science from the Department of Computer Science and Engineering, Anna University, Chennai in the year 2004. She has got more than two decades of Teaching Experience and published 50 research publications in the National/ International journals and conferences. Her research interest includes Knowledge

Representation and reasoning systems, Semantic web, Information Security and Cheminformatics.

**V. Prasanna Venkatesan,**
**Associate Professor, CDBR-SSE Lab, Department of Computer Science, Pondicherry University**
*Postal Address:* **CDBR-SSE Lab, Department of Computer Science, School of Engineering and Technology, Pondicherry University, Puducherry -14**.
Email: prasanna_v@yahoo.com
Dr. V. Prasanna Venkatesan is Associate Professor in Banking Technology Department under the School of Management at the Pondicherry University. Prior to joining the Department of Banking Technology, he was the Lecturer of Computer Science at Ramanujan School of Mathematics and Computer Science. He has published one book and more than thirty research papers in various journals, edited book volumes and conferences. He has earned B.Sc. (Physics) from Madras University and MCA, M.Tech (CSE) and PhD (CSE) from Pondicherry University, Pondicherry, India. His research interest includes Software Engineering, Object Oriented Modeling and Design, Multilingual Software Development, and Banking Technology.

## Abstract

In spite of existence of many standard security mechanisms for ensuring secure e-Commerce business, users still fall prey for online attacks. One such simple but powerful attack is 'Phishing'. Phishing is the most alarming threat in the e-Commerce world and effective anti-phishing technique is the need of the hour. This paper focuses on a novel anti-phishing browser plug-in which uses information hiding technique - Steganography. A Robust Message based Image Steganography (RMIS) algorithm has been proposed. The same has been incorporated in the form of a browser plug-in (safari) called Pixastic. Pixastic is tested in an online banking scenario and it is compared with other well-known anti-phishing plug-in methods in practice. Various parameters such as robustness, usability and its behavior on various attacks have been analysed. From experimental results, it is evident that our method Pixastic performs well compared to other anti-phishing plug-ins.

## INTRODUCTION

Internet has changed the life of human significantly and it has dominated many fields including e-Commerce, e-Healthcare etc. Internet increases the comfort of human life, on the other hand it also increases the need for security measures too. For example all web browsers and servers take almost every care to make guarantee the safe business through internet. Still they are vulnerable to attacks such as phishing. In this attack, the attacker tries to mimic as legitimate site and gather critical information from the user which in turn will be used to make control of the user's valuable and critical information.

The Anti-Phishing Working Group (APWG), an association of internet service providers is collecting information on Phishing incidents from financial institutions, online-retailers, and other IT-companies. The collected data is published in the monthly Phishing Activity Trends Report, which clearly shows the dramatically increase in Phishing attacks. Figure 1 shows the number of phishing site detected in various countries in the year 2010 by APWG. From surveys, it is observed that millions of customers are at risk of affected by Phishing in recent year. APWG further states that financial and payment services are the major sector which falls prey for phishing.

In this paper a novel Robust Message based Image Steganography (RMIS) algorithm has been proposed and the same has been incorporated in the form of a browser plug-in. The proposed method has been implemented and it is compared with other similar anti-phishing browser plug-in. The rest of the paper is organized in the following manner: section 2 deals with state of art of the anti-phishing and Steganography techniques; section 3 concentrates on the novel Robust Message based Image Steganography algorithm; section 4 focuses on Implementation and experimental set up of Pixastic browser plug-in and section 5 evaluates and compares the existing anti-phishing plug-in methods with proposed method and section 6 concludes the paper.

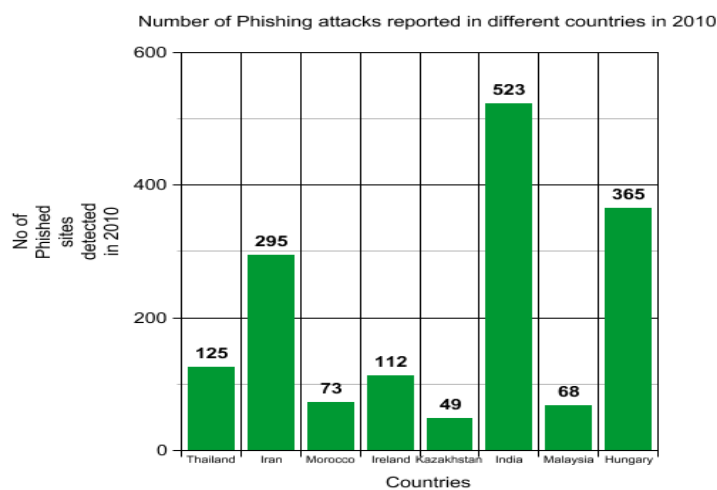

**Figure 1:** Number of Phishing sites detected in various countries

## STATE OF ART ANTI-PHISHING TECHNIQUE AND STEGANOGRAPHY

### Anti-Phishing Technique Literature Survey

Based on thorough literature survey on the available anti-phishing techniques, they have been classified broadly into four categories: Analyzing E-mails (Andre Bergholz et al., 2008; Ian Fette, Norman Sadeh and Anthony Tomasic, 2007; Wei-Chih Hsu and Tsan-Ying Yu, 2010), Website content analysis (Chinmay Somanet al., 2008; Justin Maet al., 2009; Justin Ma et al., 2009; Sujata Garera et al., 2007), Authentication methods (Rachna Dhamija and J.D. Tygar, 2005) and Plug-in methods (Pawan Prakash et al., 2010). Since our method falls in the plug-in category brief survey has been done on other anti-phishing methods such as Analyzing E-mails, Website content Analysis and Authentication method and the detailed survey has been done on anti-phishing plug-ins.

**Brief survey of Analyzing E-mails, Website content Analysis and Authentication methods**
Email is the most primary source of phishing attack. The methods (Andre Bergholz et al., 2008; Ian Fette, Norman Sadeh and Anthony Tomasic, 2007; Wei-Chih Hsu and Tsan-Ying Yu, 2010), detects phishing by analyzing the content of the e-mail. Fette et.al., identified ten different characteristic of mails to identify phishing mail. Andre Bergholz et.al., in his method uses statistical classification methods along with feature extraction technique such as Dynamic Markov Chains (DMC) and Latent Class-Topic Models (CLTOM). Wei-Chih Hsu in his method uses SVM classifier combined along with anti-spam techniques.

Anti-Phishing methods which are based on Website content analysis focus on URL and page content of the website (Chinmay Somanet al., 2008; Justin Maet al., 2009; Justin Ma et al., 2009; Sujata Garera et al., 2007). Various machine learning algorithms such as Logistic Regression (LR), Support Vector Machine (SVM), Random Forest (RF) and Neural Network were employed to differentiate the phishing site and the legitimate site.

Authentication mechanisms have also been used to address the phishing issue. Companies like Yahoo and Microsoft have their own authentication protocols that will limit spam to user's mail. Two factor authentication mechanisms is widely used by many financial organizations to prevent phishing. In (Rachna Dhamija and J.D. Tygar, 2005) methods the website will be allowed to prove the identity to the user by the shared image authentication method.

**Detailed Survey of Anti-Phishing Browser Plug-in**
The top anti-phishing plug-ins reported in literature are discussed in detail here. Netcraft plug-in was introduced in the year 2005 and any user can install this plug-in in Mozilla browser. Any site that the user access through this plug-in installed browser will display its host location and the risk rating of the site. On seeing, this information user can get to know the originality of the website. If the site is fishy user can report to the Netcraft it will validate the site and if found guilty it will be stored in the blacklisted database to prevent any further prey for that phished site.

TrustWatch is the toolbar designed especially for Internet explorer that verifies the website identity by displaying domain name and by verifying whether the URL is in the black listed database. If the URL is matched with the entry in the black listed database, it will warn the user.

Spoof Stick is a plug-in that helps user to detect fake website by displaying the domain information in the browser. Pre-requisite of this plug-in is that user should be aware of the valid domain name from where this website has been launched. Before entering their user credentials user has to ensure whether the domain name displayed by the Spoof stick is valid.

ScamBlocker identify the phished site by using eleven tips that reveals the validity of the website. Some of the tips are checking the false urgency, mails with spelling and grammar mistakes. ScamBlocker is integrated along with Earthlink mailbox, so any person who maintains account with Earthlink their incoming mails will be scanned for Phishing threat and only if it is found ignorant mail will be shown in the inbox.

PhishNet (Pawan Prakash et al., 2010) has two major components first component grows the blacklist database by generating URL variations from known Phished link. Second Component assigns a score to each URL by matching the targeted URL with the URLs generated in the first component. The limitations of the top anti-phishing plug-ins are tabulated in the Table 1.

**Table 1:** Limitations of the existing anti-phishing plug-ins

| Plug-in | Limitations |
| --- | --- |
| Netcraft | The user may not be aware of the host place of all the website being accessed. |
| TrustWatch/ PhishNet | The entered URL is checked with all the entries in the black listed database. It's a time consuming process since the blacklisted database keeps growing day by day and the possibility that the user falls prey to the newly developed phished URL until it get entered into the black listed database is very high. |
| Spoof Stick | Domain name of the website is displayed at the browser. Careful analysis of the displayed URL is necessary and it is completely depended on the user awareness of the domain name. |
| ScamBlocker | Most of the Phishing attack is done through mail. Hence the characteristic of the mail which contains the phished link has been studied. Scam blocker scans every mail for the studied pattern and based on the validity it sends to user's inbox. If the Phishers follow a new mail pattern then this method fails to detect. |

**Steganography Techniques - Literature Survey**
Steganography is one of the information hiding technique which conceals the secret message into any digital medium like image, audio, video files. The components involved in Image Steganography process is shown in figure 2. Steganography has wide range of useful applications e.g., Smart Id card (Jain. A. K and Uludag. U, 2002), Secret communication between parties (Xindiao et al., 2010), network steganography (K.Szczyporski and W.Mazurcyzk, 2011), healthcare (Der-chyuandou, Ming-chiang Hu and Jiang-Lung Liu, 2009), banking (Thiyagarajan P, Aghila G and Prasanna Venkatesan V, 2011) etc. One such area where it can be applied to give security is e-commerce. This paper uses Steganography concept in the browser plug-in to prevent phishing attack. The proposed plug-in technique uses novel Robust Message based Image Steganography algorithm.

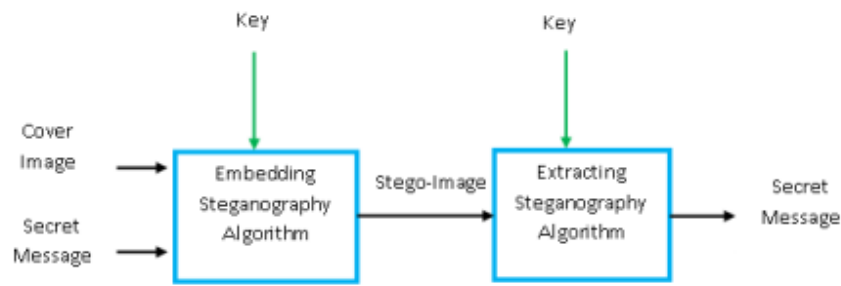

**Figure 2:** Image Steganography Process Diagram

Generally, the Steganography technique works at either in the spatial domain or the frequency domain. This work discusses a spatial domain steganography algorithm and its plug-in development. In addition, some of the famous spatial steganography algorithms are also discussed. Least Significant Bit (LSB) (Johnson. N. F. and Katzenbeisser. S. C, 2000) algorithm is the most popularly known spatial domain steganography algorithm that replaces the least significant bit of every pixel by the secret message bits. (Potdar. V.M, Han. S and Chang. E, 2005) in their method divide the cover image into small portion and embed the secret data in the all the regions so that the image withstands the cropping attack. (Shirali-Shahreza. M. H and Shirali-Shahreza. M, 2006) in their method exploited the Persian and Arabic letters punctuation marks to hide secret messages. Colour palette is also used in steganography where the LSB's of each pixel has been considered for modification based on the colour palette. The file formats .bmp and .png are the most popular choice of steganography methods though it couldn't withstand the statistical attack and compressions.

In a nut shell the above listed steganography methods suffers from the following limitations
- The secret data are embedded in all the pixels sequentially
- The embedding rate is static
- The sensitivity of the channel in which the data is to be embedded has not been ensured by proper check
- The location of the pixels for embedding the secret messages does not depend on the secret message which is to be embedded

The above limitations of the existing spatial domain Steganography techniques are addressed in our proposed Robust Message based Image Steganography (RMIS) algorithm and the limitations of the existing anti-phishing plug-in in table 1 are addressed by using the RMIS algorithm as browser plug-in.

**PIXASTIC USING ROBUST MESSAGE BASED IMAGE STEGANOGRAPHY ALGORITHM**

This section is divided into two parts in first part detailed discussion is done on Robust Message based Image Steganography (RMIS) algorithm and in the second section architecture of the Pixastic browser plug-in is discussed in detail.

**Robust Message based Image Steganography Algorithm**
The three main phases of Robust Message based Image Steganography (RMIS) technique are a) Preprocessing b) Embedding and c) Extracting.

**Preprocessing**
Preprocessing is the first step in RMIS technique. The input to the processing phase is the secret message and the output is the embedding sequence. The secret message that is to be embedded is converted into binary values. The binary values are then grouped in two bits per group and it is converted to decimal values. This sequence of number is called as 'embedding sequence'.
The embedding sequence is multiplied with image size (row*column) and the obtained value is called as 'Stego-Key'. Stego-key is converted into binary and the binary sequence is called as 'embedding rate sequence'.

**Embedding**
Embedding is the second phase of RMIS technique. Embedding phase hides the secret messages into the given cover image in such as way that the resultant stego-image is not differentiable by Human Visual System (HVS). The following steps explain the major steps in embedding process

a) Embedding sequence that is obtained from the preprocessing steps is used here to find the pixel where secret data bits are to be embedded

b) The each and every bit in the embedding sequence is mapped to a particular pixel in the cover image

c) Fixation of the indicator channel for each pixel
   If the bit in the embedding sequence is '0' then skip that pixel from embedding
   If the bit in the embedding sequence is '1' then fix the 'Red' Channel in that pixel as the indicator channel
   If the bit in the embedding sequence is '2' then fix the 'Green' Channel in that pixel as the indicator channel
   If the bit in the embedding sequence is '3' then fix the 'Blue' Channel in that pixel as the indicator channel

d) Fixation of data and third channel
   Once the indicator channel is fixed in a pixel, find the lowest channel from the remaining two channels in that pixel and fix the lowest channel as the 'data channel'
   The left out channel is named as the 'third channel'

e) With the help of 'embedding rate sequence' embed the secret message bits in the data channel after ensuring that there is no major change in the color of the channel

f) Based on number of bits embedded in the step e the Least Significant Bit of the third channel has to be changed in order to communicate how many bits embedded in the channel to the extraction part

g) The above process continues until all the secret message bits are embedded

**Extracting**
The Extraction phase extracts the secret message embedded from the stego-image using the same secret key as in embedding phase.

Following are the main steps in extraction phase
a) The stego-key is obtained from the embedding part through secure channel.

b) Once the stego-key is obtained from the counterpart, it is divided with the size of the image and the resultant number is the 'embedding sequence' which gives the idea about the pixel which contains the secret bits

c) Select the pixel which contains the secret bits and extract the bits embedded using 'embedding rate sequence'

d) Extraction is continued until all the bits in the secret message bits are extracted

The preprocessing step and the embedding phase are taken care by the bank and the extraction phase is exercised in the browser with the help of Pixastic browser plug-in at the client side.

**Architecture of Pixastic Browser Plug-in**
Plug-in is a software component that adds specific intelligence which enhances the performance or security of the software application. The architecture of the proposed anti-phishing plug-in is shown in the figure 3 below. Following two conditions are pre-requites for working of Pixastic browser plug-in
Any bank website who wishes to use Pixastic plug-in should incorporate the Stego-image generated from Robust Message based Image Steganography embedding algorithm in their website.
Users who is having internet banking facility with the above bank should install the Pixastic Plug-in from the legitimate bank website.

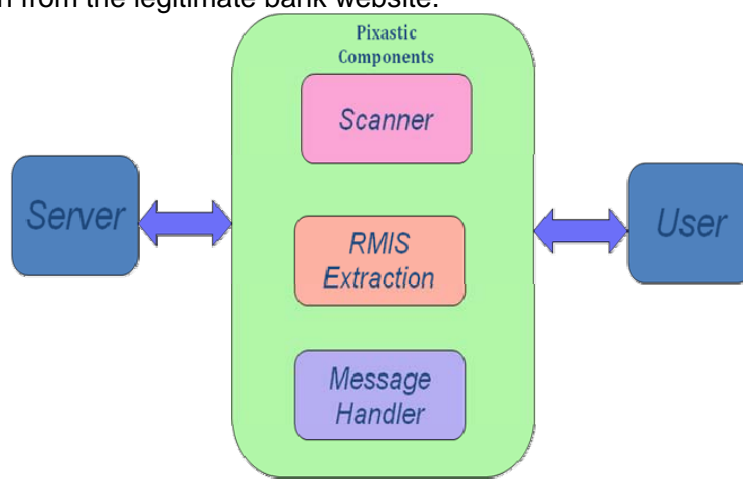

**Figure 3:** Architecture of Pixastic Browser Plug-in Pixastic components have three main components they are a) Scanner b) RMIS Extraction and c) Message Handler.

a) Scanner
Scanner scans the address bar for URL and check whether the domain name for which this plug-in was developed is there in domain name part of the URL. If the domain name or any of the sub-string of the domain name for which the plug-in was developed is found in the address bar Pixastic plug-in will be triggered.

b) RMIS Extraction
Once the Pixastic is triggered it tries to locate the stego-image in the website and it extracts the secret message using RMIS extraction algorithm.

c) Message Handler
Once the extracted secret message matches with the message in the plug-in, then the user is allowed to access the website. If there is a mismatch then the user is warned about the authenticity of the website and all the controls in the website are blocked.

The workflow of the Pixastic browser plug-in is shown in the figure 4.

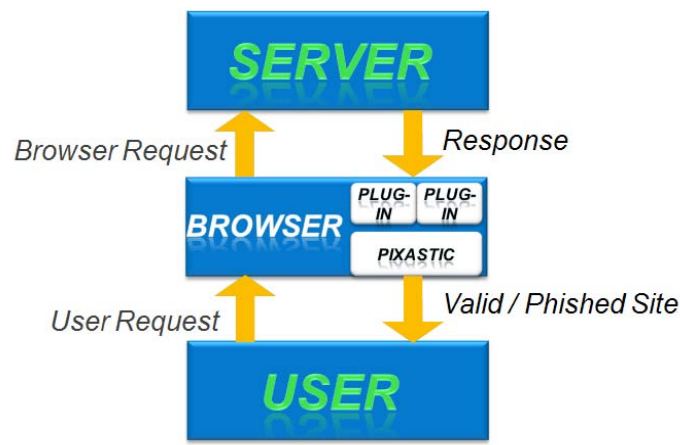

**Figure 4:** Workflow of Pixastic Browser Plug-in

### IMPLEMENTATION OF THE PIXASTIC METHOD

The plug-in development experience report (Thomas Raffetseder, Engin Kirda and Christopher Kruegel, 2007) gave us the insight about the difficulties involved in plug-in development. Tutorial for the Safari plug-in development is obtained from. The Pixastic plug-in consists of code that locates and extracts message from stego-image in the website. The sample bank website was developed for experiment purpose and it is ported in the SSE lab blade server. Five clients were connected to the blade server and in all the client machines Pixastic plug-in were installed.

For experiment, the secret message was incorporated in the bank logo. Once the web address contains the substring of the domain name in the address bar Plug-in is triggered and it tries to extract the secret message from stego-image using Robust Message based Image Steganography algorithm. Once the extracted secret message matches with the secret message in the Pixastic plug-in the user is allowed to access the website else warning is thrown regarding the website authenticity and all the controls in the webpage is blocked further so it prevents user from entering their credentials. Figure 5 shows the installation of Pixastic browser plug-in. Figure 6 depicts the working of Pixastic plug-in on legitimate website and Figure 7 shows the working of Pixastic plug-in for phished site.

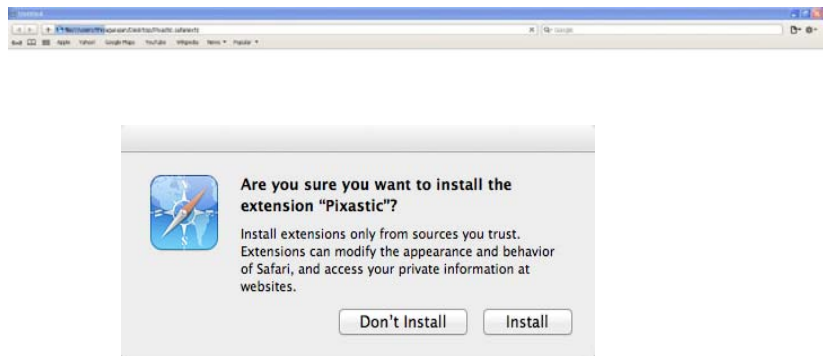

**Figure 5:** Screen shot showing the installation of Pixastic Plug-in

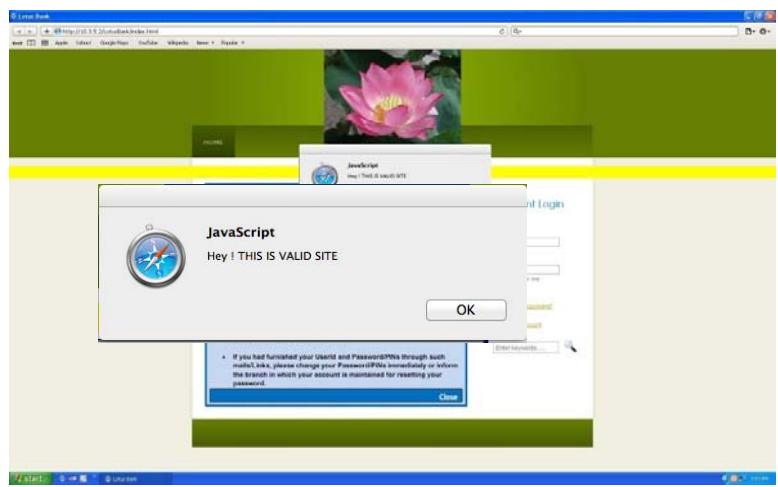

**Figure 6:** Screen shot showing the working of Pixastic on legitimate Site

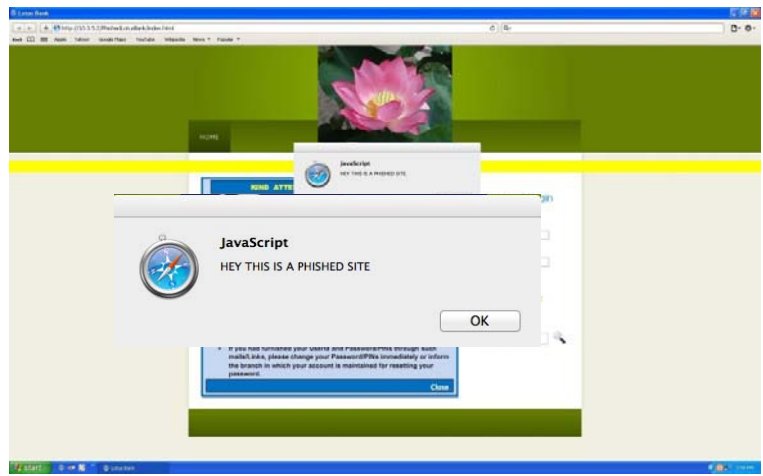

**Figure 7**: Screen shot showing the working of Pixastic on Phished Site

**EVALUATION AND COMPARISON OF THE PIXASTIC PLUG-IN**

The proposed Pixastic browser plug-in is compared and evaluated with similar plug-ins reported in the literature and their pros and cons have been analyzed. Pixastic browser plug-in was evaluated under two categories such as

a) Pixastic's resistance against known and possible attacks
b) Usability of Pixastic browser plug-in

**Pixastic's resistance against known attacks**
Various attacks on anti-phishing client plug-ins are reported in (Dinei A. F. Florêncio and Cormac Herley, 2006) such as page load attack, zero-day attack in black list database, redirection and distribution attacks. These attacks behavior on Pixastic browser plug-in is tested on these attacks and the results are shown in table 2.

**Possible attacks in Pixastic browser Plug-in**
Apart from the well known attacks the possible attacks on Pixastic plug-in was shown below
    a) Brute force attack
    b) DNS Spoofing attack
    c) Print Screen Attack

**Table 2:** Behavior of Pixastic browser plug-in in various Plug-in attacks

| S.No | Attacks on Anti-phishing Client Plug-in | Whether Pixastic Plug-in resist against attack? | Reason |
|---|---|---|---|
| 1. | Broken links and Delaying Page load Attack | No | Since the image is loaded on page load event Pixastic plug-in fails to address this attack |
| 2. | Problems in the Black list approaches | Yes | Pixastic Plug-in doesn't depend on black list database |
| 3. | Problems in the White list approaches | Yes | Pixastic Plug-in doesn't depend on white list database |
| 4. | Redirection and Distributed Attacks | Yes | Even if the phished site is originated from different place Pixastic plug-in able to detect it since it looks for the secret message inside the image in the bank website. |
| 5. | Problem of getting users to alter their behavior | Yes | Pixastic methods apart from warns user about the site validity it also disable the controls of the phished website |

**Brute force attack**

Robustness is measured by the difficulty level of the intruder to break the key for any technique. Given any RGB image intruder may try brute force attack to get back the embedded secret message. In Pixastic browser plug-in stego-key is derived from the message which is to be embedded. Apart from ensuring the dynamicity in key which is used for embedding dynamicity is also encompassed in the frequency of bits embedded and in selecting the pixel for embedding. Table 3 shows the message length and the number of distinct pattern that the attacker has to try to break the stego-key used in Pixastic browser plug-in.

**Table 3:** Length of the message and number of distinct pattern attacker has to try to break the stego-key

| Length of the Message (Bytes) | Length of key derived from message (Bits) | Number of distinct pattern attacker has to try to break the stego-key |
|---|---|---|
| 3 | 12 | 118098 |
| 7 | 28 | 5083731656658 |
| 10 | 40 | 2701703435345984178 |

In proposed method each character is converted into binary, which will consist of eight bits, these eight bits are grouped into two bits, so four groups are obtained from each character. Hence,
Length of key derived from message = length of the message * 4
Number of distinct pattern that the attacker has to try to break the stego-key = $3^{(n-2)}*2$ ways
In our experiments we have tested with the secret message of length greater than 10. Therefore it is highly difficult for attackers to crack the stego-key using brute force attack.

**DNS spoofing attack**
The role of DNS is to resolve the IP address for the web address in the URL. In DNS Spoofing attack, the web address is mapped to the IP address of the server which is not legitimate. Once the illegitimate website is loaded into the browser and if the Pixastic plug-in is installed for that website, it tries to find the validity of the website by extracting the secret message embedded. Since the website is illegitimate, Pixastic will throw a warning message and disable the controls in the website. From experiments, it is observed that Pixastic prevents DNS Spoofing attack.

**Print Screen Attack**
Since the Pixastic plug-in deals with Image Steganography attacker may try to capture the images in the website and try to extract the secret message embedded in the Stego-image from the website with all possible stego-key. This experiment is tested against our Pixastic method and the results were shown in table 4.

**Table 4:** Results tried on image taken from phished bank website with wrong Stego-key

| Embedded Secret Message in cover medium | Secret message obtained by extracting bits from all pixels in stego image |
|---|---|
| Pondicherry University | T+*m*-&#tKK`w0    -xy jg&klk |
| SSE Lab | <<]4]nD \|:8*~v&N-FKKeW;' |

Pixastic browser plug-in is tested extensively for the well-known and possible attacks and from the above results, it is obvious our method withstands these attacks.

**Usability of Pixastic browser plug-in**
Usability is the one of the main criteria for evaluating the anti-phishing browser plug-in. The Usability of the Pixastic plug-in is evaluated based on the following parameters (Li. Linfeng and Helenius Marko, 2007).

- Visibility of Result
- Prevent user from accessing Phished website
- Flexibility
- Aesthetic and Privacy
- Portability

**Visibility of Result**
The Visibility of the plug-in result should be clear. After verifying the validity of the website the result should be clear and obvious. The result should not be too technical and it should be understandable by the user who doesn't have any computer knowledge or naïve users. Pixastic browser plug-in shows very clear indication to the user if the website is phished site by displaying the authenticity of the website.

**Prevent user from accessing Phished Website**
After the website has been analyzed for phishing if the website is genuine then the website is allowed for further processing from the user side. If the website is found to be phished site then warning is displayed. But users may overlook the warning message and try to proceed with the fake website. In Pixastic to avoid this, all the controls in the fake website have been disabled which prevents the user from entering their credentials which is not done in other plug-in methods.

**Flexibility**
Financial website that incorporates this plug-in can change the secret message that is embedded inside the image or can change the image where the secret message is embedded or can change the key that is used for embedding the secret message. Pixastic plug-in is flexible to all these changes. This flexibility also enhances the dynamicity of the proposed plug-in.

**Aesthetic and Privacy**
Our Pixastic plug-in user interface is very simple and it does not add any icon in the browser. Since Pixastic plug-in depends on the image in the website the copying and the right click of the website is prevented which prevent the hacker from copying the image.

**Portability**
Currently the Pixastic browser plug-in is implemented in the safari browser. The extraction code is written in java script and it can be portable to any browser provided the browser has the compatibility of reading the pixel value.

Pixastic browser plug-in is compared with other anti-phishing browser plug-ins against various parameters and it is shown in table 5. The grading has been given to different plug-ins where +++ stands for very good, ++ for good, + for average, -- for substandard. Evaluation criteria that are followed to rate these plug-in are explained below.

**Robustness**
Behavior of DNS Spoofing attack has been analyzed here and it is found that Netcraft plug-in and Pixastic plug-in will resist this attack. In Net craft plug-in, it will only display the host place from where the DNS Spoofed website is launched. If the user does not aware of the host place of the legitimate website from where it is launched, then this method will fail. Hence + grading has been given for Netcraft. But Pixastic will resist DNS Spoofing attack since the stego-image for which the Pixastic plug-in is looking will be present only in the legitimate website, hence ++ grading has been given for Pixastic. Not all other plug-ins addressed DNS spoofing attack hence – grading has been given.

| Plug-ins | Robustness | Visibility of the Result | Prevent user from accessing Phished website | Aesthetic | Training Costs |
|---|---|---|---|---|---|
| Net craft | + | + | -- | + | -- |
| Trust Watch | -- | + | -- | + | -- |
| Phish Net | -- | + | -- | + | -- |
| Spoof Stick | -- | + | -- | + | -- |
| Scam Blocker | -- | + | -- | + | -- |
| Pixastic Browser Plug-in | ++ | +++ | +++ | ++ | +++ |

**Table 5:** Comparison of Pixastic browser plug-in with other existing Anti-Phishing Plug-ins

**Visibility of the Result**
All the plug-ins in tables 5 except Pixastic adds additional component in the browser and the validity of the website is displayed in that component. User has to look into that component portion to see the result. However, Pixastic throw the warning message to the user in message box. Therefore +++ rating was given to Pixastic and + rating was given to other plug-ins.

**Preventing user from accessing Phished site**
Pixastic apart from giving warning about phished site it will also disable all the controls in the phished site. So user's access to phished site is prevented. The rest of the plug-ins fail to prevent users from accessing phished site even if user overlooks the result hence +++ grading has been given to Pixastic and – grading was given to other plug-ins.

**Aesthetic**
Unlike other plug-ins, Pixastic does not add additional component in the browser that in turn increase the look and feel of the browser. Hence ++ rating has been given to Pixastic and – to other anti-phishing plug-ins.

**Training costs**Pixastic users may or may not have any knowledge of the domain, host place of the website that they are visiting. Therefore very good grading +++ has been given to Pixastic. Users who are using Netcraft and Spoofstick plug-in should be trained to know the host place and the domain name of the website for their safe browsing. In Trustwatch, Spoofstick and Phishnet users should be trained on the functionalities of the toolbar to prevent them from phishing. Hence – grading has been given to other plug-ins.From the above comparisons it is clear that Pixastic plug-in performs well when compared to other Plug-ins.

**CONCLUSION**

In this paper, first kind of browser plug-in Pixastic which uses Steganography is proposed and implemented. Pixastic uses Novel Robust Message based Image Steganography algorithm for embedding and extracting the secret message and it is incorporated in safari browser.  Dynamicity of RMIS algorithm is encompassed in the stego-key, embedding rate and in pixel selection. Moreover, Pixastic plug-in is quick in response since its specific for website.  Pixastic has also been compared with other anti-phishing plug-ins on parameters such as usability, behavior of plug-in on existing and possible attacks.  Pixastic   plug-in can be extended to other browsers like Mozilla, internet explorer in future.